\documentclass[preprint,12pt]{elsarticle}

\journal{Data in Brief}

\usepackage{booktabs}
\usepackage{longtable}
\usepackage{tabularx}
\usepackage{graphicx}
\usepackage{multirow}
\usepackage{hyperref}
\usepackage{pdflscape}
\usepackage{array}
\usepackage{threeparttable}
\usepackage{ragged2e}
\usepackage{amsmath}
\usepackage{subcaption}

\newcolumntype{Y}{>{\RaggedRight\arraybackslash}X}
\usepackage[T1]{fontenc}
\usepackage[utf8]{inputenc}
\usepackage{xcolor}
\usepackage{float}
\usepackage{hyperref} 
\usepackage{placeins}

\biboptions{sort&compress}

\begin{document}

\begin{frontmatter}

\title{A citizen science behavioural dataset of an urban collective-risk dilemma on air pollution mitigation under resource inequality}

% >>> Author ORDER, AFFILIATIONS, ORCIDs and the corresponding author's e-mail.
\author[openub,ubics]{Marc Sadurn\'i}
\author[openub,ubics]{Mart\'in F. D\'iaz}
\author[openub]{Juli\'an Vicens\fnref{fn1}}
\fntext[fn1]{Present address: Eurecat, Centre Tecnològic de Catalunya, Barcelona, 08005, Spain.}
\author[openub,ubics]{Anna Cigarini}
\author[openub,ubics]{Isabelle Bonhoure}
\author[openub,ubics,SH]{Miquel Montero} 
\author[openub,ubics]{Josep Perell\'o\corref{cor}}
\cortext[cor]{Corresponding author.
\ead{josep.perello@ub.edu}}
\affiliation[openub]{organization={OpenSystems, Departament de F\'isica de la Mat\`eria Condensada, Universitat de Barcelona}, city={Barcelona}, postcode={08028}, country={Spain}}
\affiliation[ubics]{organization={Universitat de Barcelona Institute of Complex Systems (UBICS)}, city={Barcelona}, country={Spain}}
\affiliation[SH]{organization={Serra H\'unter Fellow, Generalitat de Catalunya},
            state={Catalonia},
            country={Spain}}

\begin{abstract}
This article describes a behavioural dataset from a collective-risk dilemma  public experiment carried out in Barcelona, Spain, within the citizen science \textit{xAire} air quality project. Participants played a ten-round game in groups of six, each deciding every round how many monetary units to contribute from a private endowment toward a shared target; groups that fell short forfeited all remaining capital. Three endowment treatments held the group total fixed: an equal condition and two unequal conditions in which low-endowment participants formed either the minority or the majority of the group. Data were collected at two public outdoor venues using a tablet-based interface under lab-in-the-field protocols. Participants were recruited on-site from a demographically heterogeneous pool of individuals passing through the venues, mostly local residents. The dataset comprises 462 participants across 77 six-player games, 4\,620 contribution decisions in total. Each participant record links the full ten-round contribution trajectory to initial endowment, wealth treatment, venue, district of residence and its measured nitrogen-dioxide concentration, and socio-demographic and attitudinal survey items covering gender, age, education, economic and working status, air-quality perception, fairness norms, environmental-justice beliefs, residential mobility, and community participation. The release provides raw and cleaned participant tables, a variable codebook, the survey instrument, the district-pollution linkage, and the preprocessing code. The data can be reused to characterise cooperative behaviour under resource inequality, benchmark behavioural and computational models of collective action, examine associations between contribution decisions and socio-demographic or environmental covariates, and compare lab-in-the-field experimental designs with conventional human-behaviour laboratory settings. 
\end{abstract}


\end{frontmatter}

% =====================================================================
% VALUE OF THE DATA
% =====================================================================
\section*{Value of the Data}
\begin{itemize}
  \item The dataset records 4{\,}620 individual contribution decisions from a controlled collective-risk dilemma (CRD) with three calibrated endowment treatments, placing it among one of the largest participant samples in the experimental CRD literature on climate mitigation \cite{milinski2008collective,tavoni2011inequality,burtonchellew2013combined,brown2017avoiding,waichman2021inequality}.
  \item Each participant's complete ten-round contribution trajectory is linked to socio-demographic variables, attitudinal measures, and the measured NO$_2$ concentration of their district of residence. Participants were recruited on-site from a demographically heterogeneous pool of individuals passing through the venues, mostly local residents  and directly concerned with local air-pollution issues, providing a rare opportunity to study cooperative behaviour in a population personally affected by the environmental problem under investigation \cite{sagarra2016citizen}.
  \item The two unequal treatments hold the aggregate group endowment, collective target, and high-to-low endowment ratio (2:1) fixed, while varying whether low-endowment players form the minority or the majority of the group. This enables reuse for examining how the composition of resource inequality shapes collective behaviour independently of the magnitude of the rich-to-poor endowment ratio.
  \item The complete survey instrument, the district-level pollution linkage, and the deterministic preprocessing code are released alongside the data, so the analytical subset can be regenerated from the raw tables and the survey items reused as covariates or modelling features.
\end{itemize}

% =====================================================================
% BACKGROUND
% =====================================================================
\section{Background}
The collective-risk dilemma (CRD) is a threshold public-goods game introduced to capture the all-or-nothing structure of cooperative climate action: a group must accumulate contributions to a fixed target within a finite horizon or face a collective loss \cite{milinski2008collective}. In each round, players decide how much of their own endowment to contribute to a common fund. Contributions reduce the resources retained by the individual, whereas unspent resources remain in the player's account. At the end of the game, participants keep the portion of their endowment that was not contributed to the collective fund, creating a direct tension between individual retention and collective success. The game has been extended to unequal endowments and to varied risk and communication conditions \cite{milinski2011rich,tavoni2011inequality,burtonchellew2013combined,brown2017avoiding,vasconcelos2014climate,waichman2021inequality}. Theoretical work has additionally used the CRD framework to examine the evolutionary dynamics and timing of contribution strategies \cite{abouchakra2012evolutionary}, while collective-risk experiments have also investigated how exposure to collective risk shapes cooperative social norms \cite{szekely2021evidence}. A citizen science implementation brought the design into public settings with heterogeneous participants \cite{vicens2018resource,perello2012linking,sagarra2016citizen,perello2022new,perello2024socialphysics}, allowing social dilemmas to address a wide range of social issues, including community mental health \cite{cigarini2018mentalhealth} and gender-biased interactions in public spaces \cite{cigarini2020gender}.

\begin{figure}[tb]
\centering
\includegraphics[width=0.99\textwidth]{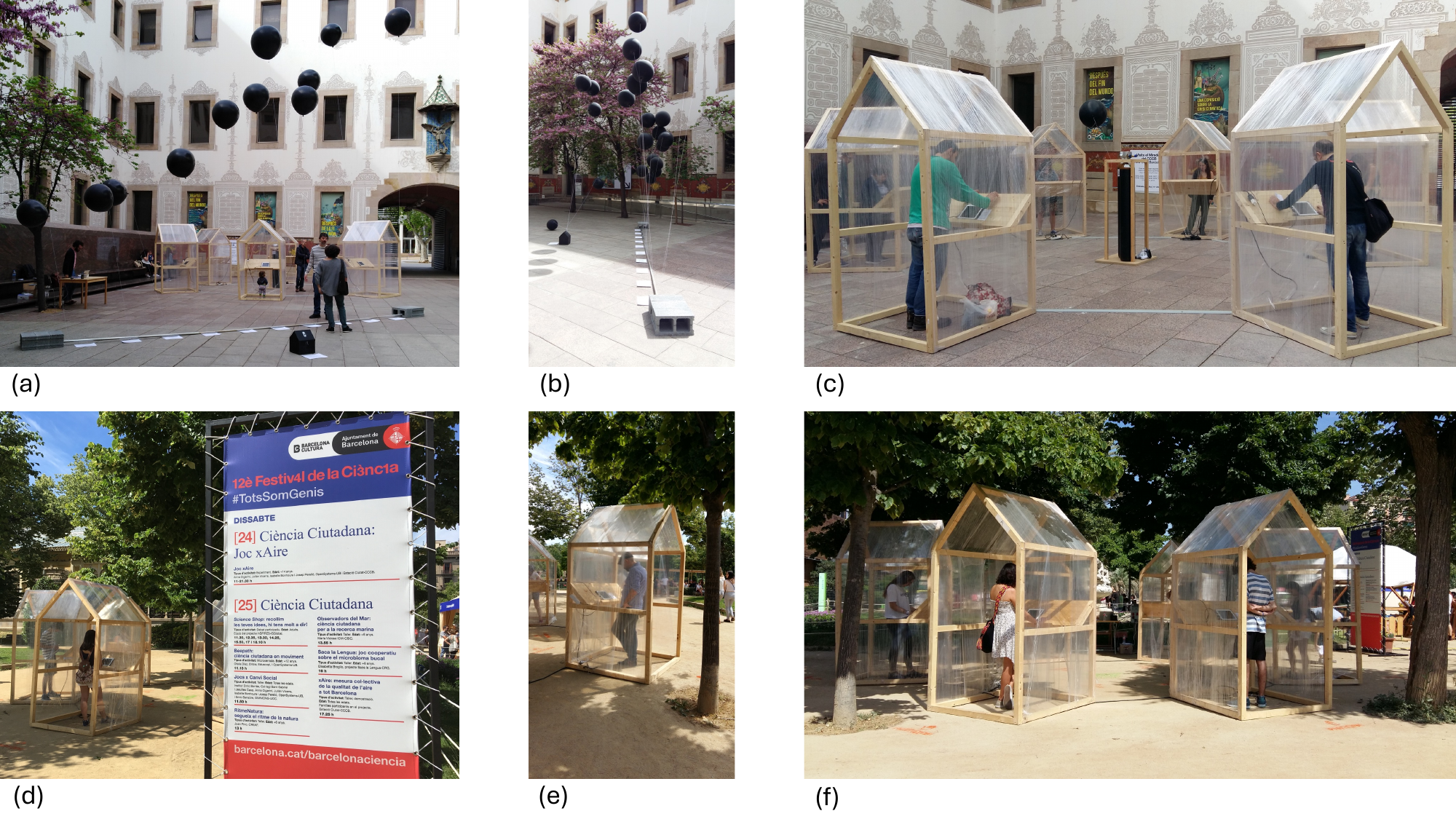}
\caption{Public outdoor settings of the \textit{xAire} Collective Risk Dilemma experiment. (a) Experimental setup at the Centre de Cultura Contemporània de Barcelona (CCCB), in the Pati de les Dones, including the six participant stations, the table hosting the local server and network connecting the six tablets, and visual elements communicating district-level air pollution data collected through the \textit{xAire} citizen science campaign. (b) Detail of the district-level air-quality information displayed in the Pati de les Dones. (c) Circular arrangement of the six stations with participants during an experimental session at the Pati de les Dones. (d) Experimental setup in Parc de la Ciutadella during the Barcelona Science Festival. (e) Close-up of one participant station, designed to prevent verbal and visual communication among participants, in Parc de la Ciutadella. (f) Arrangement of the participant stations in Parc de la Ciutadella. Setting designed by OpenSystems \cite{OpenSystems} and Domestic Data Streamers \cite{domestic}.
}
\label{fig:setting}
\end{figure}

The data described here were generated to extend this line of research within the citizen science \textit{xAire} air-quality monitoring campaign \cite{perello2021largescale,perello2021large,perello2021data}, using the Citizen Social Lab platform \cite{vicens2018citizen} and lab-in-the-field protocols for collective behavioural experiments in public settings \cite{sagarra2016citizen,perello2024socialphysics,perello2012linking}. The \textit{xAire} air-quality measurement campaign produced high-resolution measurements of nitrogen dioxide (NO$_2$) concentrations across Barcelona with more than 2\,000 participants involved and attracted considerable media attention due to growing concerns about air pollution and its health impacts. These measurements enabled the identification of neighbourhoods, and more broadly districts, with different levels of NO$_2$ exposure and informed the assignment of participants to positions associated with different resource endowments in the CRD experiment. By grounding the CRD in a locally relevant environmental problem, the experimental design transformed a widely recognized urban challenge into a collective decision-making scenario. Experimental sessions were conducted at two public venues in central Barcelona, Parc de la Ciutadella and the Centre de Cultura Contemporània (CCCB) - Pati de les Dones (Fig. \ref{fig:setting}). Recruitment followed lab-in-the-field and pop-up protocols aimed at engaging the general public rather than conventional laboratory subject pools \cite{sagarra2016citizen,perello2024socialphysics,perello2012linking}. By embedding the experiment within major public events (Sant Jordi festivity and Barcelona Science Festival), the study reached a diverse population in natural urban settings and facilitated participation by residents directly exposed to the environmental issues addressed by the \textit{xAire} campaign. The public setting of the experiment further allowed participants to continue the broader societal discussion initiated by the \textit{xAire} air-quality measurement campaign and to reflect on potential collective actions to address air pollution. In this way, resource inequality in the game was connected to a real-world environmental issue, creating an experimental framework in which the interplay between environmental inequality, cooperation, and collective action could be studied under controlled yet socially meaningful conditions.

The dataset is released to make participant-level contribution trajectories, survey responses, and contextual environmental information openly available for reuse, comparative analyses, and model benchmarking.

% =====================================================================
% DATA DESCRIPTION
% =====================================================================
\section{Data Description}
The data are released in five groups: the behavioural core (per-round contributions), the socio-demographic block, the environmental linkage (district-level pollution), the survey instrument (pre-game, comprehension, and post-game items), and a set of derived round-level indicators. Table~\ref{tab:files} lists the file inventory; Table~\ref{tab:composition} summarises the sample composition; Table~\ref{tab:summary} provides summary statistics by wealth treatment; Table~\ref{tab:no2} gives the district-level nitrogen dioxide (NO\textsubscript{2}) linkage; Table~\ref{tab:dict} is the variable dictionary for the behavioural and socio-demographic variables; Table~\ref{tab:survey} documents the full survey instrument.

\subsection{File inventory}
The released files are listed in Table~\ref{tab:files}. The integrated participant file is the primary file for reuse: it carries one row per participant with the full behavioural trace, the socio-demographic block, the survey responses, and the assigned NO\textsubscript{2} value. 

\begin{table}[htbp]
\centering
\caption{File inventory of the released dataset.}
\label{tab:files}
\begin{tabularx}{\textwidth}{@{} l Y @{}}
\toprule
File & Contents \\
\midrule
\texttt{raw/users\_xaire.csv} & Participant-level records as logged by the interface (direct identifiers removed; see Ethics Statement). \\
\texttt{raw/games\_xaire.csv} & Game-level records as logged by the interface. \\
\texttt{raw/translations\_xAire.xlsx} & Interface translation source, archived as received. \\
\texttt{processed/participants.csv} & Cleaned, integrated participant file (six-player games only): behavioural trace, socio-demographics, survey responses, NO\textsubscript{2}. Primary file for reuse. \\
\texttt{processed/games.csv} & Cleaned game-level file (six-player games only). \\
\texttt{derived/Observables\_Data.csv} & Round-level aggregate indicators by endowment level. \\
\texttt{raw/pollution.csv} & District-to-NO\textsubscript{2} linkage table. \\
\texttt{codebook/codebook.csv} & Full variable dictionary for every file. \\
\texttt{codebook/response\_codes.csv} & Code-to-label mapping (Catalan, Spanish, English) for every coded demographic and survey variable in the dataset. \\
\texttt{scripts/preprocess.py} & Deterministic transformation from \texttt{raw/} to \texttt{processed/}. \\
\texttt{scripts/codebook.py} & Regenerates \texttt{codebook.csv}. \\
\bottomrule
\end{tabularx}
\end{table}

\subsection{Sample composition}
The experiment comprised three wealth treatments: Equal ($\mathrm{E}$), in which all six participants received 40~monetary units (MU); Unequal-L ($\mathrm{L}$), in which two participants received 24 MU and four received 48 MU; and Unequal-H ($\mathrm{H}$), in which four participants received 30 MU and two received 60 MU. Thus, the $\mathrm{L}$ and $\mathrm{H}$ treatment labels distinguish whether low-endowment participants formed the minority or the majority of the group, respectively. The experimental design and treatment-assignment procedure are described in detail in Section~\ref{sec:methods}. The partition of participants and games across treatments, endowment levels, and venues is given in Table~\ref{tab:composition}.

\begin{table}[htbp]
\centering
\caption{Sample composition by wealth treatment and venue. Participant counts equal six times the number of games (only six-player games are retained). Endowments are in monetary units (MU). The experiment was performed in two locations (Parc de la Ciutadella and Centre de Cultura Contempor\`ania de Barcelona-CCCB).}
\label{tab:composition}
\resizebox{\textwidth}{!}{%
\begin{tabular}{l l c c c c}
\toprule
Treatment & Endowment per player (MU) & Games & Participants & Ciutadella & CCCB \\
\midrule
Equal      & six players $\times$ 40           & 6  & 36  & 2  & 4  \\
Unequal-L  & two $\times$ 24, four $\times$ 48 & 27 & 162 & 20 & 7  \\
Unequal-H  & four $\times$ 30, two $\times$ 60 & 44 & 264 & 26 & 18 \\
\midrule
Total      & ---                               & 77 & 462 & 48 & 29 \\
\bottomrule
\end{tabular}
}
\end{table}

\subsection{Behavioural core}
The behavioural core is the set of per-round contribution decisions. For each participant $i$ the dataset records the assigned initial endowment $e_{0,i}^\omega\in\{24,30,40,48,60\}$~MU, conditional on the wealth treatment $\omega\in\{\mathrm{E},\mathrm{L},\mathrm{H}\}$, and the ten contributions, each taking a value in $\{0,2,4\}$~MU. From these, the total contributed to the common fund, the remaining endowment, and a binary indicator of whether the participant's group reached the 120~MU target are provided. Of the 77 games, 74 reached the target and 3 did not.

\subsection{Aggregate contribution dynamics}
\begin{figure}[tb]
\centering
\includegraphics[width=0.7\textwidth]{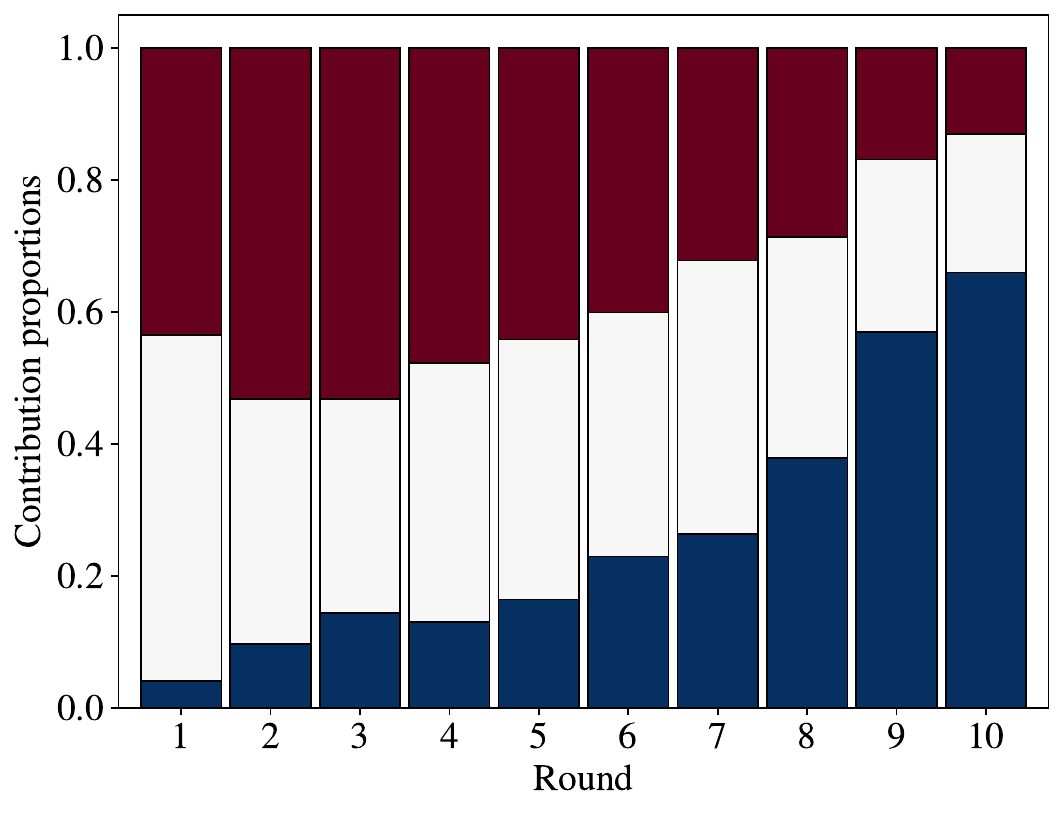}
\caption{Round-wise stacked proportions of the three possible per-round contribution choices (0, 2, and 4 MU), aggregated across all successful games and endowment classes. Within each bar, segments are ordered from bottom to top as 0 MU (blue), 2 MU (white), and 4 MU (burgundy).}
\label{fig:contrib}
\end{figure}

\begin{figure}[tb]
\centering
\includegraphics[width=0.7\textwidth]{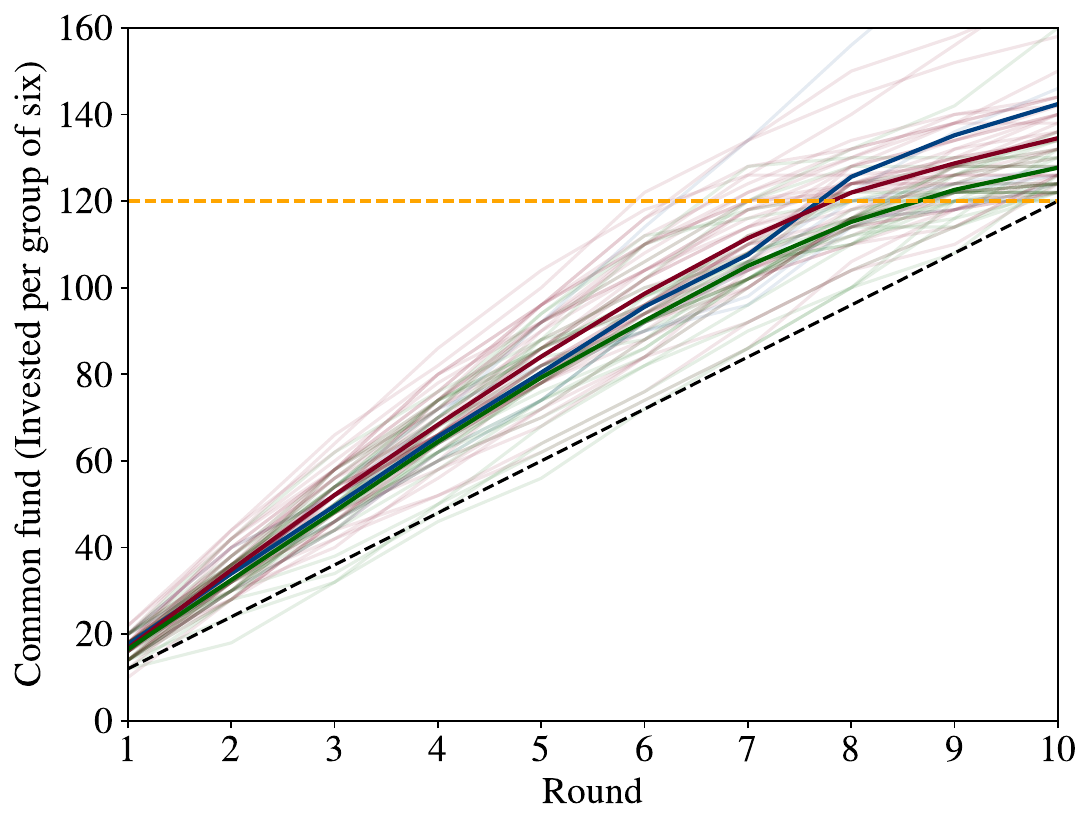}
\caption{Trajectory of the cumulative common fund (invested per group of six) over the ten rounds, across successful games. Thin lines show individual games; thick lines show the mean per wealth treatment; the horizontal orange dashed line marks the 120~MU collective target, while the black dashed line denotes the fairness benchmark (12 MU per round), i.e., the constant-rate trajectory required to reach the 120 MU target exactly at the last round. Colours correspond to the three experimental treatments, which appear in ascending order in round 10: Unequal-L (green), Unequal-H (dark red), and Equal (blue), see Table \ref{tab:composition}.}
\label{fig:fund}
\end{figure}

\begin{figure}[tb]
\centering
\includegraphics[width=0.7\textwidth]{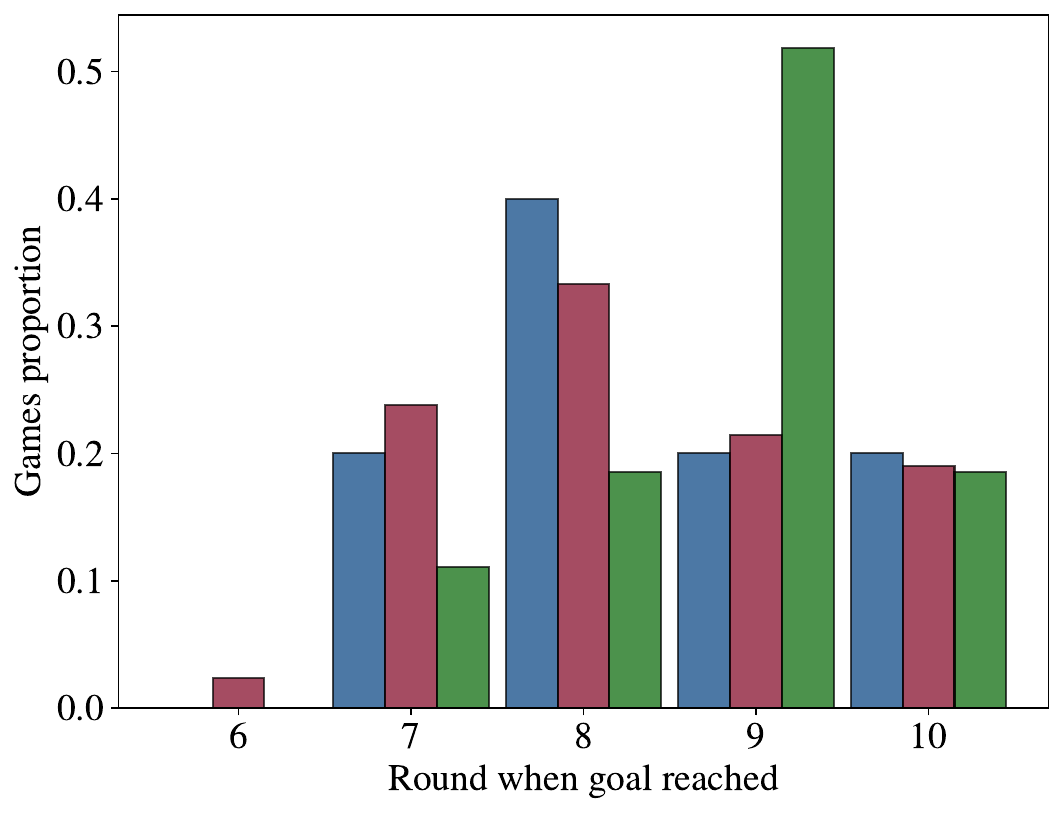}
\caption{Distribution of the round at which successful games first reached the 120~MU collective target, by wealth treatment. Colours correspond to the three experimental treatments, following the same coding as in Figure~\ref{fig:fund}. Within each round, bars are arranged from left to right as Equal (blue), Unequal-H (dark red), and Unequal-L (green), see Table \ref{tab:composition}.}
\label{fig:goalround}
\end{figure}

Figures~\ref{fig:contrib}--\ref{fig:goalround} characterise the contribution data at the aggregate level; they are descriptive depictions of the released variables and are restricted to successful games (those reaching the 120~MU target). Figure~\ref{fig:contrib} shows, for each of the ten rounds, the stacked proportion of the three possible contribution choices (0, 2, and 4~MU) across all successful games. The visible rise in zero contributions toward rounds 9--10 in Figure~\ref{fig:contrib} should not be read as decaying cooperation without qualification: because the cumulative fund is shown to players each round (Section~\ref{sec:methods}), most successful games have already reached the 120~MU target by round 8--9 (Figure~\ref{fig:goalround}), after which further contribution has no effect on the collective outcome. The late-round zero surge is therefore largely mechanical, reflecting games where the goal has already been secured, rather than a behavioural decline in willingness to cooperate. Figure~\ref{fig:fund} shows the trajectory of the cumulative common fund over rounds, with one thin line per game, thick lines denoting the mean per wealth treatment, horizontal line marking the 120~MU target, and the diagonal dashed line represents the constant-rate trajectory required to reach the 120 MU target exactly at the end of round 10 (12 MU accumulated per round). Figure~\ref{fig:goalround} shows the distribution of the round at which successful games first reached the target, separated by wealth treatment. Table~\ref{tab:summary} complements these figures by summarising game-level outcomes across wealth treatments, including the success rate, mean round of target attainment, mean total contribution, and initial and final within-game Gini coefficients \cite{owid-what-is-the-gini-coefficient}.

\begin{table}[tbp]
\centering
\caption{Summary statistics by wealth treatment for the 77 released games. Gini coefficients are calculated from the within-game endowment distribution, restricted to the games that reached the 120~MU target (74 of 77: 5 Equal, 27 Unequal-L, 42 Unequal-H), and averaged across games within each treatment. Initial Gini is based on initial endowment and final Gini on endowment remaining at the end of the game.}
\label{tab:summary}
\resizebox{\textwidth}{!}{%
\begin{tabular}{l c c c c c c}
\toprule
Treatment & Games & \begin{minipage}[c]{0.15\linewidth} Success rate (\%)\end{minipage} & \begin{minipage}[c]{0.18\linewidth}
Mean round of attainment
\end{minipage}  &  \begin{minipage}[c]{0.22\linewidth}
Mean total contribution (MU) \end{minipage}  & Initial Gini & Final Gini \\
\midrule
Equal     & 6  & 83.3  & 8.40 & 23.00 & 0.000 & 0.279 \\
Unequal-L & 27 & 100.0 & 8.78 & 21.28 & 0.133 & 0.291 \\
Unequal-H & 44 & 95.5  & 8.31 & 22.28 & 0.167 & 0.405 \\
\bottomrule
\end{tabular}
}
\end{table}

\subsection{Environmental linkage (district-level pollution)}
The environmental context of the experiment was grounded in the ten administrative districts of Barcelona, which served both as familiar territorial units for participants and as the basis for defining experimental positions. The \textit{xAire} citizen science air monitoring campaign generated high-resolution NO$_2$ measurements for each district, allowing differences in environmental exposure to be incorporated directly into the game design. These districts, which represent the primary administrative divisions of Barcelona and each account for approximately 6\% to 16\% of the city's population, correspond to well-known places of daily life for local residents. Consequently, the experimental framing connected resource inequality and collective action to a familiar local reality. Participants could therefore relate the game not only to the broader problem of air pollution in Barcelona but also to differences in environmental exposure across districts and neighbourhoods with which they were personally familiar. 

The district of residence was linked to the nitrogen-dioxide (NO\textsubscript{2}) concentration measured during the \textit{xAire} campaign \cite{perello2021largescale}. The resulting linkage (Table~\ref{tab:no2}) reports, for each district, the measured NO\textsubscript{2} concentration in micrograms per cubic metre together with a qualitative air-quality rating. The assigned concentration is also stored directly in each participant record (\texttt{no2\_level}), ranging from 39.05 to 52.38~$\mu$g\,m$^{-3}$ across districts. The Poor/Regular rating is based on the local air-quality classification applied at the time of the study, using a threshold of 40~$\mu$g\,m$^{-3}$ for annual mean NO$_2$ concentrations. Nevertheless, all district values exceed the WHO 2021 annual mean guideline of 10~$\mu$g\,m$^{-3}$ for NO\textsubscript{2} \cite{who2021aqg}. The category Out of Barcelona includes participants residing outside the city of Barcelona, with NO\textsubscript{2} values averaged across all samples collected during the \textit{xAire} campaign. The variable \texttt{no2\_level} denotes the high/low pollution-level split used to assign wealth treatments based on whether the average pollution level of the participant's district of residence was below or above the campaign average.

\begin{table}[htbp]
\centering
\caption{Residence-code to district mapping and the measured nitrogen-dioxide concentration assigned to each district in the \textit{xAire} campaign. District names follow standard Catalan spelling. "Air quality" applies a $40$~$\mu$g/m$^3$ Poor/Regular cutoff from WHO guideline category at that time (see the caveat in Section~\ref{sec:methods}). “Out of Barcelona” denotes participants residing outside the city of Barcelona; the reported NO$_2$ value corresponds to the average across all samples collected during the \textit{xAire} campaign. “Level” refers to the high/low pollution-level split used to assign wealth treatments, distinguishing districts with pollution levels below or above the campaign average (Section~\ref{sec:methods}). The level variable is displayed as High and Low for clarity; these categories are encoded as H and L, respectively, in the released \texttt{pollution.csv} dataset.}
\label{tab:no2}
\resizebox{0.8\textwidth}{!}{%
\begin{tabular}{l l c c c}
\toprule
Code & District & NO$_2$ ($\mu$g/m$^{3}$) & Air quality & Level \\
\midrule
r1 & Ciutat Vella & 50.31 & Poor & $\text{High}$ \\ r2 & L'Eixample & 52.38 & Poor & $\text{High}$ \\ r3 & Sants-Montju\"{\i}c & 46.89 & Poor & $\text{High}$ \\ r4 & Les Corts & 44.60 & Poor & $\text{Low}$ \\ r5 & Sarri\`a-Sant Gervasi & 41.88 & Poor & $\text{Low}$ \\ r6 & Gr\`acia & 43.80 & Poor & $\text{Low}$ \\ r7 & Horta-Guinard\'o & 39.53 & Regular & $\text{Low}$ \\
r8 & Nou Barris & 40.88 & Poor & $\text{Low}$ \\ r9 & Sant Andreu & 44.17 & Poor & $\text{Low}$ \\ r10 & Sant Mart\'i & 39.05 & Regular & $\text{Low}$ \\ r11 & Out of Barcelona & 46.23 & Poor & $\text{High}$ \\
\bottomrule
\end{tabular}
}
\end{table}

\subsection{Socio-demographic block}
\begin{figure}[tb]
\centering
\includegraphics[width=\textwidth]{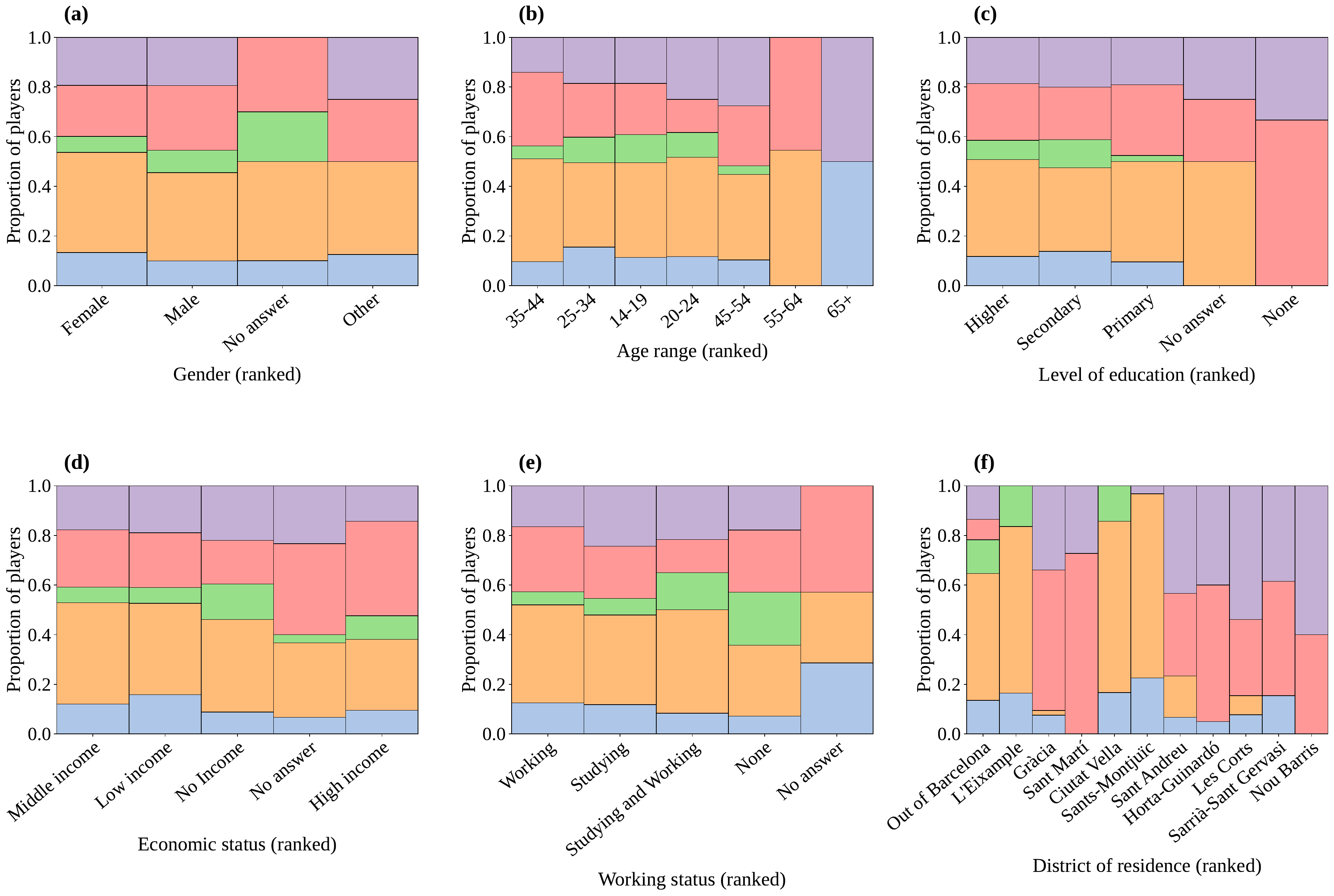}
\caption{Participant composition across socio-demographic characteristics. Proportion of participants by (a) gender, (b) age range, (c) educational level, (d) economic status, (e) working status, and (f) District of residence, ranked by participation. Stacked bar segments indicate the distribution of initial endowments: 24 MU (blue), 30 MU (orange), 40 MU (green), 48 MU (pink), and 60 MU (purple), ordered from bottom to top.}
\label{fig:demographics}
\end{figure}

Each participant reported gender, age range, educational level, economic status, working status, and district of residence. The categorical levels and their English labels are given in Table~\ref{tab:dict}; the verbatim item wording in three languages is in Table~\ref{tab:survey}. Participation spanned all ten Barcelona districts as well as out-of-city residents, and covered the full range of age bands, educational levels, and self-declared economic statuses, yielding a heterogeneous participant pool. Figure~\ref{fig:demographics} depicts the composition of the sample across these characteristics, with each category disaggregated by initial endowment.

% ----  DEMOGRAPHICS ----

\subsection{Survey instrument}
Three pre-game items captured air-quality perception, concern, and stated willingness to act collectively; three comprehension-check items verified that participants understood the game rules before play; and eleven post-game items captured beliefs about citizen science collaboration, collective efficacy, fairness norms, conditional cooperation, environmental justice, residential mobility, community participation, and air-quality salience. Table~\ref{tab:survey} documents every item with its full set of answer options. The final post-game item was an open free-text field; it has been removed from the public release to protect participant privacy (see the Ethics Statement).

\subsection{Derived round-level indicators}
The derived file provides, for each endowment level, round-level aggregates including the initial and final Gini coefficients of the endowment distribution, the mean round at which the target was reached, the proportion of games reaching the target, mean per-round contributions, and the proportion of capital contributed. These quantities are computed deterministically from the processed tables by the released code and are provided for convenience.

% =====================================================================
% TABLE: VARIABLE DICTIONARY
% =====================================================================
%\FloatBarrier
{\footnotesize
\begin{longtable}{>{\raggedright\arraybackslash}p{2.9cm}
                  >{\raggedright\arraybackslash}p{1.8cm}
                  >{\raggedright\arraybackslash}p{3.5cm}
                  >{\raggedright\arraybackslash}p{4.9cm}}
\caption{Variable dictionary for the integrated participant file (\texttt{processed/participants.csv}). The game-level and derived files are documented in \texttt{codebook.csv} in the repository. The response codes mapping for every socio-demographic variable are in \texttt{codebook/response\_codes.csv}.}
\label{tab:dict}\\
\toprule
Variable & Type (unit) & Values / range & Definition \\
\midrule
\endfirsthead
\multicolumn{4}{l}{\emph{Table~\ref{tab:dict} continued}}\\
\toprule
Variable & Type (unit) & Values / range & Definition \\
\midrule
\endhead
\midrule
\multicolumn{4}{r}{\emph{continued on next page}}\\
\endfoot
\bottomrule
\endlastfoot
\texttt{id} & Integer & unique & Participant identifier; join key for the participant tables (replaces the original nickname). \\
\texttt{partida\_id} & Integer & unique & Game identifier; groups the six players of a game. \\
\texttt{control\_wealth} & Categorical & EQUAL; UNEQUAL-L; UNEQUAL-H & Wealth treatment of the game (Section~\ref{sec:methods}). \\
\texttt{endowment\_initial} & Integer (MU) & 24; 30; 40; 48; 60 & Initial endowment at the start of round~1. \\
\texttt{endowment\_current} & Integer (MU) & $\geq 0$ & Endowment remaining at the end of the game. \\
\texttt{contributed\_\allowbreak public\_\allowbreak goods} & Integer (MU) & $\geq 0$ & Total contributed to the common fund over the ten rounds. \\
\texttt{winnings\_\allowbreak public\_\allowbreak goods} & Integer (MU) & $\geq 0$ & Capital retained: the remaining endowment if the group reached the target, else~0. \\
\texttt{goal\_reached} & Binary & 0; 1 & Whether the participant's group reached the 120~MU target. \\
\texttt{R1}--\texttt{R10} & Integer (MU) & 0; 2; 4 & Per-round contribution in rounds 1 to 10 (ten columns). \\
\texttt{no2\_level} & Numeric ($\mu$g/m$^{3}$) & 39.05--52.38 & District NO$_2$ concentration, assigned from residence (Table~\ref{tab:no2}). \\
\texttt{gender} & Categorical & Male; Female; Other; No answer & Self-reported gender identity. \\
\texttt{age\_range} & Categorical & 14--19; 20--24; 25--34; 35--44; 45--54; 55--64; 65+ & Self-reported age band. \\
\texttt{economic\_status} & Categorical & No income; Low income; Middle income; High income; No answer & Self-reported economic status. \\
\texttt{educational\_level} & Categorical & None; Primary; Secondary; Higher; No answer & Highest completed educational level. \\
\texttt{working\_status} & Categorical & None; Studying; Working; Studying and Working; No answer & Self-reported working status. \\
\texttt{residence} & Categorical & 10 Barcelona districts and Out of Barcelona (see Table~\ref{tab:no2}) & Self-reported district of residence. \\
\texttt{frame\_pr1}--\texttt{pr3}, \texttt{verification\_pr1}--\texttt{pr3}, \texttt{enquesta\_final\_pr1}--\texttt{pr10} & Categorical / ordinal & see Table~\ref{tab:survey} & Pre-game, comprehension, and post-game survey items. \texttt{verification\_pr1}--\texttt{pr3} are the three comprehension-check items (Table~\ref{tab:survey}) that confirmed a participant understood the contribution rule, the group target, and the failure condition before play began; the interface required a correct answer to proceed, so all three are constant in the released data (zero-variance by construction, not a data-quality issue -- reusers should treat them as a design/comprehension record for the sample, not as a substantive predictor).\texttt{enquesta\_final\_pr11} (free text) is withheld for privacy. \\
\end{longtable}
}

% =====================================================================
% TABLE: SURVEY INSTRUMENT
% =====================================================================
{\footnotesize
\begin{longtable}{>{\raggedright\arraybackslash}p{3.2cm}
                  >{\raggedright\arraybackslash}p{7.4cm}
                  >{\raggedright\arraybackslash}p{3.0cm}}
\caption{Survey instrument: pre-game, comprehension, and post-game items with answer options (English). The verbatim trilingual wording (Catalan, Spanish, English) and the response codes mapping for every coded variable are in \texttt{codebook/response\_codes.csv}. The underlying source translations are archived as received in \texttt{raw/translations\_xAire.xlsx}.}
\label{tab:survey}\\
\toprule
Item & Question & Answer options \\
\midrule
\endfirsthead
\multicolumn{3}{l}{\emph{Table~\ref{tab:survey} continued}}\\
\toprule
Item & Question & Answer options \\
\midrule
\endhead
\midrule
\multicolumn{3}{r}{\emph{continued on next page}}\\
\endfoot
\bottomrule
\endlastfoot
\multicolumn{3}{l}{\emph{Pre-game framing}}\\
\texttt{frame\_pr1} & How would you rate the quality of the air you breathe where you live? & Bad; Normal; Good.\\
\texttt{frame\_pr2} & Are you concerned about the air quality where you live? & Little; Somewhat; A lot. \\
\texttt{frame\_pr3} & Would you perform a collective action to improve air quality in Barcelona? & Yes; No. \\
\addlinespace
\multicolumn{3}{l}{\emph{Comprehension checks (constant; gated to correct answer)}}\\
\texttt{verification\_pr1} & In a group of six, if everyone contributes equally, the minimum each must contribute to reach the target. & 10; 20; 30; 40 MU. \\
\texttt{verification\_pr2} & Whether a group total of 100 coins over ten rounds reaches the target. & Yes; No; I don't know. \\
\texttt{verification\_pr3} & Whether failing to collect 120 coins means losing all personally saved coins. & Yes; No; I don't know. \\
\addlinespace
\multicolumn{3}{l}{\emph{Post-game}}\\
\texttt{enquesta\_final\_pr1} & What do you think might be the major benefit of the collaboration between citizens and scientists? & Opening my horizons and changing my perspective; Doing better science in academic contexts; Contributing to improvements in our environment; Making knowledge more public and accessible; I think there are no benefits. \\
\texttt{enquesta\_final\_pr2} & If data and evidences are gathered collectively, do you believe that it is possible to propose changes and improvements in a more effective way? & Totally disagree; Disagree; Neither agree nor disagree; Agree; Totally agree. \\
\texttt{enquesta\_final\_pr3} & At the beginning of the game, did you expect to reach the common goal? & Yes; No; I don't know. \\
\texttt{enquesta\_final\_pr4} & In the game, the contributions of each participant should be proportional to their capital: the richest should contribute more and the poorest should contribute less. & Totally disagree; Disagree; Neither agree nor disagree; Agree; Totally agree. \\
\texttt{enquesta\_final\_pr5} & In the game, if others contribute little I should also contribute little. If others contribute a lot I should also contribute a lot. & Totally disagree; Disagree; Neither agree nor disagree; Agree; Totally agree. \\
\texttt{enquesta\_final\_pr6} & We are all equally exposed to air pollution. & Totally disagree; Disagree; Neither agree nor disagree; Agree; Totally agree. \\
\texttt{enquesta\_final\_pr7} & Have you ever considered living in another district where air quality is better? & Yes; Sometimes; No. \\
\texttt{enquesta\_final\_pr8} & Do you participate in your neighbourhood’s community life? & Never; Rarely; Sometimes; Often; Always. \\
\texttt{enquesta\_final\_pr9} & How important do you think it is the issue of air pollution in Barcelona compared to the social emergencies that the city is facing? & Way less important; Less important; Equally important; More important; Way more important. \\
\texttt{enquesta\_final\_pr10} & Are you aware of actions or interventions aimed at improving air quality in Barcelona? & Not at all; Slightly; Somewhat; Moderately; Extremely. \\
\texttt{enquesta\_final\_pr11} & Ideas for the city council to improve air quality. & Free text (withheld; privacy). \\
\end{longtable}
}

% =====================================================================
% EXPERIMENTAL DESIGN, MATERIALS AND METHODS
% =====================================================================
\section{Experimental Design, Materials and Methods}\label{sec:methods}

\subsection{The \textit{xAire} project, venues and recruitment}
The experimental data were collected in two lab-in-the-field sessions organised by the OpenSystems group of the Universitat de Barcelona within the \textit{xAire} project, a large-scale citizen science air-quality campaign in which communities around twenty Barcelona primary schools measured nitrogen-dioxide concentrations across the city \cite{perello2021largescale}. Sessions took place between April and June 2018 at two public outdoor venues: Parc de la Ciutadella and the Centre de Cultura Contempor\`ania de Barcelona (CCCB, Pati de les Dones). Recruitment followed  protocols designed to engage the general public rather than a pre-registered subject pool, broadening demographic coverage relative to conventional laboratory samples \cite{sagarra2016citizen,perello2024socialphysics}. In Parc de la Ciutadella, a major urban park in central Barcelona characterised by extensive green areas, cultural institutions, and high pedestrian flow, the experiment was conducted within the framework of the Barcelona Science Festival (June 9 and 10), which featured multiple simultaneous public activities across the park. At the CCCB, the experiment took place in the Pati de les Dones, an open courtyard within a centrally located cultural complex in a densely frequented urban area, during the Sant Jordi festivity (April 23), when large numbers of people gather informally in the surrounding streets (Fig. \ref{fig:setting}). In both settings, these contexts enabled the recruitment of participants from all districts of Barcelona (71.2\%) as well as from outside the city but mostly belonged to Barcelona metropolitan area which shared same air-quality issues (28.8\%) and were exposed to similar urban air-quality challenges, thus preserving the local relevance of the experimental setting. After data cleaning and preprocessing, the final dataset comprises 48 valid games (288 participants) from Ciutadella and 29 games (174 participants) from CCCB.

\subsection{Game mechanics}
Participants played a stylised collective-risk dilemma \cite{milinski2008collective,vicens2018resource}. Each game involved $N=6$ players over $R=10$ rounds. In round $k \in \{1, \dots, R\}$, player $i\in \{1, \dots, N\}$ chose a contribution $c^{\,\omega}_{k,i}\in\{0,2,4\}$~MU from their remaining endowment, where $\omega$ indexes the wealth treatment. The collective target was fixed at half the total initial endowment,
\begin{equation}
T=\tfrac{1}{2}\sum_{i=1}^{N} e_{0,i}^{\,\omega}=120~\text{MU},
\end{equation}
identical across treatments because every treatment summed to $\sum_{i} e^{\,\omega}_{0,i}=240$~MU. The cumulative common fund and the evolving individual endowment are
\begin{equation}
\phi^{\,\omega}_{k}=\sum_{k'=1}^{k}\sum_{i=1}^{N} c^{\,\omega}_{k',i},
\qquad
e^{\,\omega}_{k,i}=e^{\,\omega}_{0,i}-\sum_{k'=1}^{k} c^{\,\omega}_{k',i}.
\end{equation}
At the end of round $R$ the payoff was binary: if $\phi^{\,\omega}_{R}\ge T$ the group succeeded and each player retained their remaining endowment; otherwise all players retained nothing. Importantly, participants are fully informed about these consequences prior to starting the experiment. 

After each round, participants observed the contributions of the other group members, the total amount accumulated in the common fund, and their own remaining endowment (see Fig. \ref{fig:screenshots}). This feedback enabled participants to update their decisions based on both collective progress and their individual position. The choice set in round $k$ is also implicitly censored by the endowment remaining after rounds $1,\dots,k-1$: the interface only offers a contribution in $\{0,2,4\}$~MU if that amount does not exceed the player's remaining endowment, so a player who has already spent their full endowment can no longer select a positive contribution. This holds without exception across the released data: in all 462 participant records, every per-round contribution is less than or equal to the endowment remaining at that point in the game. Reusers computing round-level feasible choice sets (e.g. a "budget-constrained" indicator alongside Table~\ref{tab:dict}) should apply this rule rather than treat $\{0,2,4\}$ as unconditionally available in every round.

\subsection{Wealth treatments}
Initial endowments were assigned under one of three treatments while holding the group total fixed at 240~MU:
\begin{equation}
e^{\,\omega}_{0,i}=
\begin{cases}
40~\text{MU}, & \omega=\mathrm{E},\ \forall i,\\[2pt]
24~\text{MU}, & \omega=\mathrm{L},\ i\le 2,\\
48~\text{MU}, & \omega=\mathrm{L},\ i>2,\\[2pt]
30~\text{MU}, & \omega=\mathrm{H},\ i\le 4,\\
60~\text{MU}, & \omega=\mathrm{H},\ i>4.
\end{cases}
\label{eq:initialendowments}
\end{equation}
The $\mathrm{L}/\mathrm{H}$ label names the size of the low-endowment group, not the severity of inequality, and this convention should be stated explicitly since a reader could plausibly parse $\mathrm{H}$ as "high inequality": in Unequal-L, low-endowment players form the minority (two of six); in Unequal-H they form the majority (four of six). Both unequal treatments hold the rich-to-poor endowment ratio fixed at 2:1 (48:24 and 60:30) and vary only the \emph{composition} of the group, not the ratio. The wealth treatment of each game is recorded in the \texttt{control\_wealth} field.

These three treatments capture variations in resource distribution that reflect stylised socio-economic divisions observed in urban environments, ranging from fully egalitarian to structurally imbalanced setups where wealthier or poorer individuals form the majority. By varying the structure of initial capital while holding the group size and game mechanics across treatments, the design enables a comparative analysis of how resource heterogeneity influences individual behaviour and group outcomes in collective-risk scenarios.

\subsection{Treatment assignment procedure}
Treatment assignment was not based on uniform randomisation, but on the NO$_2$ pollution categories associated with participants' districts of residence. District-level NO$_2$ concentrations were obtained from the \textit{xAire} campaign data for Barcelona \cite{perello2021data} and classified into high- and low-pollution categories. Each participant was therefore assigned a pollution level according to their district of residence (Table~\ref{tab:no2}).

For each six-player game, let $N_\text{High}$ and $N_\text{Low}$ denote the number of participants classified in the high- and low-pollution categories, respectively. Treatment assignment followed the deterministic rule:
\begin{equation}
    \omega =
    \begin{cases}
        \mathrm{E}, & N_\text{High} = 0 \ \text{or}\ N_\text{Low} = 0,\\
        \mathrm{H}, & N_\text{High} > N_\text{Low},\\
        \mathrm{L}, & N_\text{Low} > N_\text{High}.
    \end{cases}
\end{equation}
Thus, when one pollution category constituted the majority of the group, the corresponding unequal treatment was assigned: Unequal-H when $N_\text{High}$ participants were the majority and Unequal-L when $N_\text{Low}$ participants were the majority. If all six participants belonged to the same pollution category, the Equal treatment was assigned.

In the special case of an equal split, $N_\text{High}=N_\text{Low}=3$, the mean residential NO$_2$ concentration of the six participants,
\begin{equation}
    \overline{\mathrm{NO}_2}
    =
    \frac{1}{6}\sum_{i=1}^{6}\mathrm{NO}_{2,i},
\end{equation}
was compared with the reference Barcelona value of $46.23\,\mu\mathrm{g/m^3}$ stored in the experimental platform. A mean equal to this reference produced the Equal treatment, a lower mean produced the Unequal-L treatment, and a higher mean produced the Unequal-H treatment.

Once the treatment had been determined, participants were ordered from highest to lowest residential NO$_2$ concentration before the corresponding endowment vector was assigned. Consequently, in the unequal treatments, participants associated with the highest residential pollution levels received the lower initial endowments: 24 MU in Unequal-L and 30 MU in Unequal-H. The complete implementation of this deterministic assignment procedure is publicly available in the released experimental code \cite{githubstem}. This data-driven procedure ensured that wealth distributions within each game reflected realistic socio-environmental inequalities present in Barcelona’s urban landscape. Because treatment assignment depended on the residential pollution profile of the six participants assembled for each game rather than on random allocation, treatment and venue are not independent (see Table~\ref{tab:composition}); this should be taken into account when reusing the data for treatment-wise comparisons.

\subsection{Interface, feedback and incentive}
\begin{figure}[tb]
\centering
\includegraphics[scale=0.73]{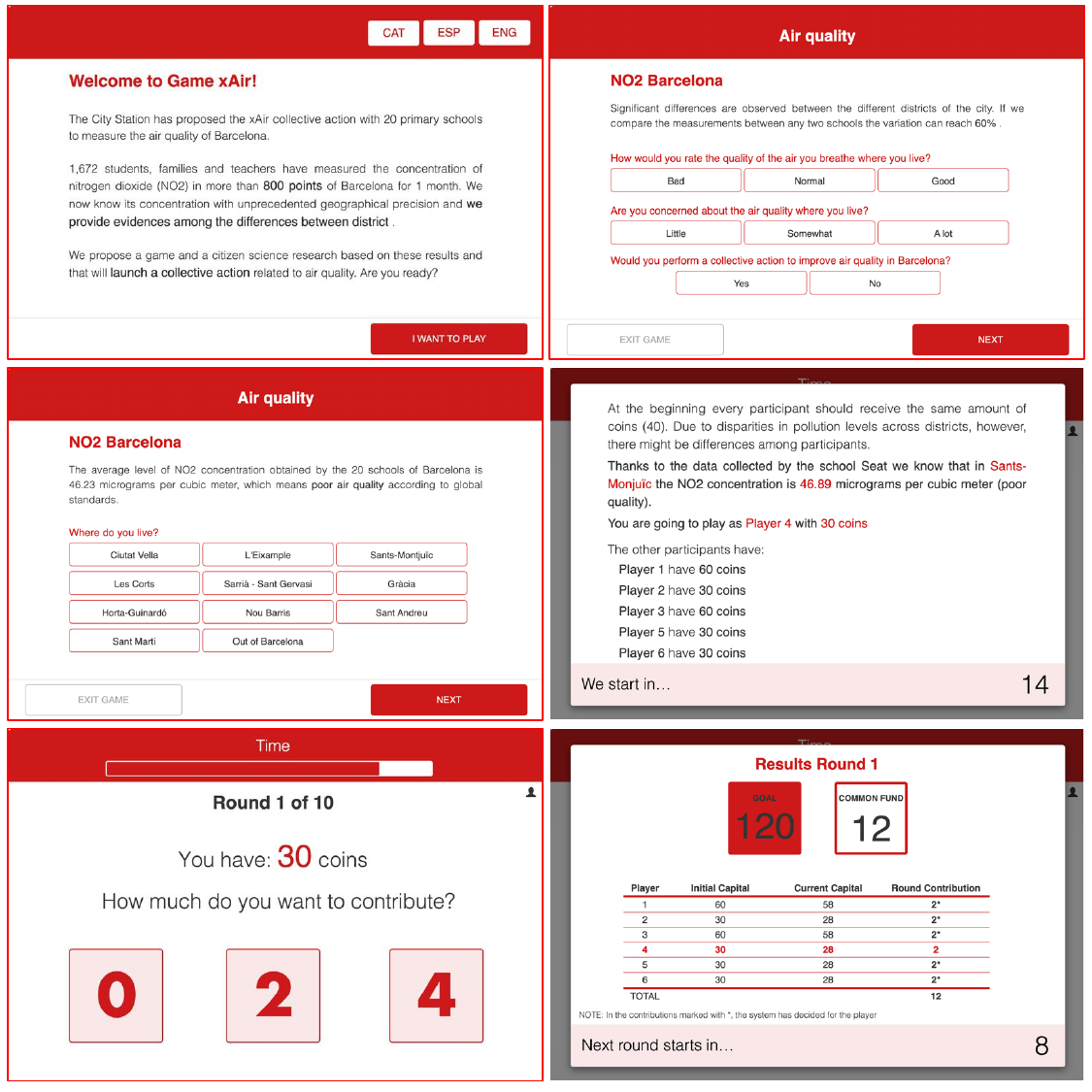}
\caption{Selection of screenshots from the tablet-based interface used in the public experiments. From left to right and top to bottom, the sequence shows the welcome screen; contextualization and self-perception of air quality; contextualization and residence-district reporting; initial experiment instructions; the decision screen used in each round; and the feedback screen presented after each round.}
\label{fig:screenshots}
\end{figure}

All sessions used a tablet-based interface built on the Citizen Social Lab platform \cite{vicens2018citizen}, with trained facilitators delivering standardised instructions; communication between participants was not permitted. At the outset, clear instructions were provided with the support of trained facilitators to ensure full comprehension of the task and its implications. During each round $k$, participants faced a countdown timer limiting the time available (15 seconds) to decide whether to contribute 0, 2, or 4 MU to a common fund. While participants were not explicitly informed about the method of endowment and treatment assignment, the interface visibly displayed each player's initial capital. After each round, every participant was shown three pieces of feedback: the cumulative total in the common fund, the amount contributed by each participant in the previous round, and each participant’s initial and current endowment. This made inequality immediately apparent, introducing an implicit fairness dilemma: whether wealthier individuals should contribute more, or whether less-endowed players should compensate for limited group effort. Participants took approximately 10-15 minutes to complete the experiment, including the welcome phase, pre-game framing, comprehension checks, the CRD experiment itself, and the surveys administered before and after the game. Figure \ref{fig:screenshots} presents a selection of screenshots from the tablet-based interface, showing the sequence of screens encountered by participants during the experiment.

Beyond the collective target of promoting air-quality mitigation actions with public schools of the city, a private incentive was also in place. At the beginning of the session, participants were informed that they would receive bookshop vouchers, with the final amount proportional to the fraction of their initial endowment retained after the ten rounds. Eligibility for the voucher payment was conditional on the group reaching the collective target. Participants received monetary rewards in the form of 1 EUR vouchers according to their performance in the experiment. In total, 859 vouchers were distributed to 496 participants, corresponding to an average reward of 1.73 vouchers per participant (standard deviation  = 0.82; median = 2; range: 1-5 vouchers).

\subsection{Pre-game framing, comprehension checks and surveys}
Before play, participants completed a short pre-game block on air-quality perception, concern, and willingness to act collectively, followed by three comprehension-check items confirming that they understood the contribution rule, the failure condition, and the target. After play, participants completed an eleven-item post-game block covering beliefs about collaboration, collective efficacy, fairness norms, conditional cooperation, environmental justice, residential mobility, community participation, and air-quality salience. All items, answer options, and their trilingual wording (Catalan, Spanish and English) are provided in the repository and summarised in Table~\ref{tab:survey}.

\subsection{Data preprocessing}
A first consistency-cleaning pass, conducted by J.~Vicens for the original data release \cite{vicens2018resource}, removed (i) participants who did not complete the full sequence of decisions and (ii) registrants who never began play. After that pass, the working dataset comprised 83 games across the two venues (52 at Ciutadella and 31 at CCCB), with 496 human participants and games of varying composition.

Some games in the raw dataset also include automated players (“robots”), which were used when a complete group of human participants was not available. At each round, a robot selected uniformly at random among the contribution options available to it through the game interface and therefore did not follow a strategic or adaptive policy. Robots are identified in \texttt{raw/users\_xaire.csv} by the binary variable \texttt{is\_robot}, which takes value 1 for automated players and 0 for human participants.

To ensure comparability of group dynamics across treatments, only games with exactly six human players were retained; this removed 7 games and yielded 77 valid games (48 from Ciutadella, 29 from CCCB) and 462 participants, each contributing once per round over ten rounds for 4{\,}620 recorded decisions. The treatment partition is 6 Equal (36 participants), 27 Unequal-L (162 participants), and 44 Unequal-H (264 participants). By endowment level the participant counts are 24~MU: 54; 30~MU: 176; 40~MU: 36; 48~MU: 108; and 60~MU: 88. The deterministic code that reproduces the cleaned subset from the raw tables is provided in \texttt{scripts/}.

\subsection{Data validation and integrity checks}

Three comprehension-check items (Table~\ref{tab:survey}) verified that participants understood the contribution rule, the collective target, and the failure condition before the game began. The experimental interface required a correct answer to each item before participants were allowed to proceed, ensuring that all participants started the game with a minimum understanding of its objectives and consequences.

The processed dataset was subjected to a set of internal consistency checks before release.
Across all 462 participant records, the sum of the ten recorded round contributions equals the total amount contributed to the common fund,
\begin{equation}
    \sum_{k=1}^{10}  \texttt{Rk} = \texttt{contributed\_public\_goods},
\end{equation}
and the individual accounting identity
\begin{equation}
    \texttt{endowment\_initial}
    =
    \texttt{endowment\_current}
    +
    \texttt{contributed\_public\_goods}
\end{equation}
holds for every participant. The retained payoff is likewise consistent with the collective outcome,
\begin{equation}
    \texttt{winnings\_public\_goods}
    =
    \texttt{endowment\_current}
    \times
    \texttt{goal\_reached}.
\end{equation}

In addition, all per-round contributions satisfy the feasible-choice constraint imposed by the participant's remaining endowment at that point in the game. At the game level, each retained game contains exactly six human participants, and the value of \texttt{goal\_reached} is consistent across all six members of the same game. These checks were performed during preprocessing prior to the public release of the dataset.
% =====================================================================
% LIMITATIONS
% =====================================================================
\section*{Limitations}
The dataset has several limitations. The Equal treatment is under-represented (6 games, 36 participants) relative to the two unequal treatments, limiting balanced treatment-wise comparisons. All sessions were conducted in a single city and at two venues, so the participant pool, while demographically heterogeneous, is geographically specific. Although participants shared a common local context and were broadly familiar with the environmental challenges represented in the experiment, recruitment took place in public venues and participation was voluntary, introducing self-selection relative to a pre-registered sample. From a citizen science perspective, this feature can also be considered a strength, as the dataset captures the behaviour and attitudes of concerned and engaged citizens interacting with a real local environmental issue. Furthermore, treatment assignment was determined by the pollution-category composition of participants assembled in each game rather than by individual randomisation. These features enhance the ecological validity of the dataset but should be considered when making causal treatment comparisons. Treatment and venue are not statistically independent, and analyses comparing treatments should account for possible venue or session effects. The game-success outcome is highly skewed toward success (74 of 77 games reached the target), leaving few unsuccessful games for outcome-conditioned reuse. The district-level NO\textsubscript{2} value is an area-level attribute assigned by district of residence, not an individual exposure measurement. The data were collected in 2018, and Barcelona's NO$_2$ levels and air-quality policies have evolved since then; accordingly, the attitudinal responses should be interpreted as reflecting the environmental and policy context at the time of data collection. Finally, the open free-text survey item and the participant-chosen identifier have been withheld from the public release to protect privacy and are therefore not available for reuse.

% =====================================================================
% ETHICS STATEMENT
% =====================================================================
\section*{Ethics Statement}
The authors confirm that the work was carried out in accordance with the Declaration of Helsinki. The Universitat de Barcelona Ethics Committee (IRB00003099) approved the experiment. All participants read and signed the informed consent form and parental/legal guardian consent if appropriate. No privacy issues were observed to be in conflict with the public release of the underlying processed data. A copy of the informed-consent text (Catalan, Spanish, and English) is provided in the repository.

% =====================================================================
% CREDIT
% =====================================================================
\section*{CRediT Author Statement}
% >>> ACTION: confirm with all co-authors (trio-consistent with Cities and MethodsX).
\textbf{Marc Sadurn\'i:} Data curation, Formal analysis, Software, Validation, Visualization, Writing -- original draft, Writing -- review \& editing. \textbf{Mart\'in F. D\'iaz:} Data curation, Validation, Visualization, Writing --original draft, Writing -- review \& editing. \textbf{Juli\'an Vicens:} Conceptualization, Data curation (original collection), Investigation, Methodology, Software, Supervision, Validation, Writing -- review \& editing. \textbf{Anna Cigarini:} Conceptualization, Investigation, Data curation (original collection). \textbf{Isabelle Bonhoure:} Conceptualization, Investigation, Project administration, Resources. \textbf{Miquel Montero:} Supervision, Funding acquisition, Investigation, Methodology, Project administration, Writing -- review \& editing. \textbf{Josep Perell\'o:} Conceptualization, Supervision, Funding acquisition, Investigation, Methodology, Project administration, Resources, Writing -- review \& editing.
% =====================================================================
% DECLARATION OF COMPETING INTEREST  (mandatory)
% =====================================================================
\section*{Declaration of Competing Interest}
The authors declare that they have no known competing financial interests or personal relationships that could have appeared to influence the work reported in this article.

% =====================================================================
% ACKNOWLEDGEMENTS
% =====================================================================
\section*{Acknowledgements}
The initiative was conceived as part of City Station \cite{cccbcity}, a temporary environmental health platform designed to promote collaborative research and public engagement with science. This platform was embedded within the “After the End of the World” exhibition at the Centre de Cultura Contemporània de Barcelona (CCCB) \cite{cccbafter}, highlighting the intersection between science, culture, and civic action. The \textit{monitoring} campaign for air quality \textit{xAire} was made possible with the support of DKV  and 4sfera Innova \cite{esfera}, and in collaboration with Mobile Week Barcelona \cite{mobileword}, the Barcelona Institute of Global Health (ISGlobal) \cite{ISGlobal}, Mapping for Change \cite{mappingforchange} and OpenSystems of the University of Barcelona \cite{OpenSystems}. We thank all the volunteers for their participation. We also thank the collaboration of RUN Design, Domestic Data Streamers \cite{domestic}, Centre de Cultura Contempor\`ania de Barcelona, and Barcelona City Council. We also thank the collaboration of Anna Broll, Mario Corea, Diana Escobar, Guillermo Espada, Ferran Espanyol, Nadala Fern\'andez, Miquel Nogu\'es and Elisenda Poch. The study was partially supported by MICIU/AEI/ 10.13039/501100011033, grant number PID2019-106811GB-C33 [JV, AC, IB, MM and JP]; by MICIU/AEI/10.13039/501100011033 and by ``ESF Investing in your future'', grant number PRE2020-093266 [MS]; by MICIU/AEI/ 10.13039/501100011033 and by “ERDF/EU”, grant numbers PID2022-140757NB-I00 and PID2025-171705NB-I00 [MS, IB, MFD, MM and JP]. We also acknowledge the support of Generalitat de Catalunya through Complexity Lab Barcelona, grant number 2021SGR00856 [MS, IB, MFD, MM and JP].

% =====================================================================
% DATA ACCESSIBILITY
% =====================================================================
\section*{Data Accessibility}
The data described in this article are openly available in CORA. Repositori de Dades de Recerca (CORA.RDR), the Dataverse-based repository operated by the Consorci de Serveis Universitaris de Catalunya (CSUC), at \url{https://doi.org/10.34810/DATA3672} \cite{vicens2018xairedata}.

\section*{Declaration of Generative AI and AI-assisted technologies in the writing process}
During the preparation of this work the authors used Claude (Anthropic) in order to improve the language, grammar, syntax, and clarity of the manuscript. All scientific content, methodological decisions, analyses, and conclusions were conceived and written primarily by the authors. After using this tool, the authors reviewed and edited the content as needed and take full responsibility for the content of the publication. 

\bibliographystyle{elsarticle-num}
\bibliography{references}

\end{document}